\documentstyle[fleqn,12pt,epsf]{article}
\renewcommand{\baselinestretch}{1.5}

\def\gord{$ \raisebox{-.3ex}{$\stackrel{>}{_{\sim}}$} $}
\def\lord{$ \raisebox{-.3ex}{$\stackrel{<}{_{\sim}}$} $}

\begin{document}
\begin{flushright}
NUC-MIN-96/7-T; SUNY-NTG-96-24
\end{flushright}
\begin{center}
{\Large \bf {Strangeness and Metastable Neutron Stars:\protect\\
What Might have Happened to SN1987A }}
\end{center}
\vskip 1cm
\noindent {\small Immediately after they are born, neutron stars are 
characterized by an entropy per baryon of order unity and by the presence of
trapped neutrinos. If  the only hadrons in the star are nucleons, these effects
slightly reduce the maximum mass relative to cold, catalyzed matter.  However,
if negatively charged particles in the  form of hyperons, a kaon condensate, 
or quarks are also present, these effects  result in an increase in the maximum
mass of $\sim 0.2{\rm M}_{\odot}$ compared to that of a 
cold, neutrino-free star. This
could lead to the delayed formation of a  black hole; such a scenario is
consistent with our present knowledge of  SN1987A.}                     
\vskip.75cm
\noindent {\large{1. A NEUTRON STAR IS BORN}}
\vskip.75cm

\noindent Neutron stars are thought to originate from the core collapse of
stars of mass $\gord8$ solar masses (M$_{\odot}$) at the end of their lives. 
The collapsing core subdivides into an inner region, in which the
infall velocity is proportional to the radius, and an outer region, 
which collapses less quickly.  When the central density of the inner
core significantly exceeds that of an atomic nucleus, the core bounces.   The
resulting collision with the still-infalling outer core generates a  shock
wave.   Although the shock wave rapidly moves outwards, it stalls within the
outer core at a radius of a few hundred km.  Accretion of infalling
matter through the shock generates sufficient energy in the form of neutrinos
to support the shock and prevent a recollapse for up to several tenths of a
second.  Present calculations \cite{snth} suggest that convection
supplies additional energy which results in a supernova explosion and the 
expulsion of most of the mass exterior to the core.
                              
Initially, the remnant's gravitational mass is less than 1 M$_\odot$.  
The remnant, termed
the protoneutron star, is lepton rich and has an entropy per baryon 
of $S\simeq1$ (in units of Boltzmann's constant $k_B$).   
The leptons include both electrons and neutrinos, the latter being trapped in
the star because their mean free paths in the dense matter are of order 1 cm,
whereas the stellar radius is about 15 km.  Accretion onto the neutron star
increases its  mass to the 1.3--1.5 M$_\odot$ range, and should mostly cease
after a second. It then takes about 10--15 s \cite{birth} for the trapped
neutrinos to diffuse out, and in the diffusion process they leave behind most
of their energy, heating the protoneutron star to fairly  uniform entropy
values of about $S=2$.   Cooling continues as thermally-produced neutrinos
diffuse out and are emitted.  After about 50 s, the star becomes completely 
transparent to neutrinos and the neutrino luminosity drops precipitously
\cite{bur}.   
                       
The fate of the dense remnant left behind from the explosion depends upon the
equation of state (EOS) and the amount of additional material that falls back
onto it.  General relativity implies that there is a maximum mass for a given
equation of state.  Observations of PSR 1913+16 \cite{Taylor} and causality
\cite{rr} limit the maximum mass of cold, catalyzed matter $M_{max}$ to the
range 1.443 M$_\odot \lord M_{max}\lord 3$ M$_\odot$.  However, the maximum
mass of matter with abundant trapped leptons, $M_{max}^L$, may be larger or
smaller than $M_{max}$.  There are, therefore, two possible ways that a black
hole could form after a supernova explosion.  First, accretion of sufficient
material could increase the remnant's mass to a greater value than either
$M_{max}$ or $M_{max}^L$ and produce a black hole, which then appears on the
accretion time scale \cite{bbw}.  Second, if accretion is insignificant after a
few seconds, and if $M_{max}^L>M>M_{max}$, where $M$ is the final remnant mass,
then a black hole will form as the neutrinos diffuse out \cite{prep,tpl,
keiljan,pcl,subside}, on the deleptonization time scale of 10--15 s.  This 
scenario, involving metastable neutron stars, is the topic of this
Comment; a recent review is to be found in Ref. \cite{prep}.
         
The existence of metastable neutron stars has some interesting implications. 
First, it could explain why no neutron star is
readily apparent in the remnant of SN 1987A despite our knowledge that one
existed until at least 12 s after the supernova's explosion.  Second, it
would suggest that a significant population of relatively low mass black holes
exists \cite{bb94}, one of which could be the compact object in the 
X-ray binary 4U1700-37 \cite{bww}.

\vskip.75cm
\noindent {\large{2. STRUCTURE OF NEWBORN NEUTRON STARS}}
\vskip.75cm

\noindent How is the stellar structure, particularly the  maximum mass,
influenced by the trapped neutrinos? (The finite entropy, as we shall see,
plays a lesser role.) In order to investigate this  question, one needs the 
EOS up to $\sim10$ times the baryon density encountered in the center of a
nucleus.  The EOS at such high densities is not known with any certainty. 
Nevertheless, recent work \cite{glenhyp,kapnel,gleqk} has emphasized the
possibility that  hyperons, a condensate of $K^-$ mesons, or $u$, $d$, and $s$
quarks,  may be present in addition to  nucleons and leptons.   These
additional components can appear separately or in combination with one
another.  Notice that all of these cases introduce the strange $s$ quark into
the neutron star and involve negatively-charged,  strongly-interacting
particles. Compared to a star containing just plain-vanilla nucleons and
leptons, the presence of these additional  components qualitatively changes 
the way in which the  structure of the star depends upon neutrino trapping
\cite{prep}. In particular, they permit the existence of metastable neutron
stars, which could collapse to black holes during deleptonization.     
                                                                   
The composition of the star is determined by two important physical
constraints.  The first is charge neutrality -- there must be equal numbers of
positively and negatively charged particles at a given density.
The second is beta equilibrium -- because the time scales of weak
interactions, including those of strangeness violating processes, 
are short compared to the dynamical time scales of evolution, chemical 
equilibrium is achieved among the various possible constituents. 
For example, the process $p+e^- \leftrightarrow n+\nu_e$ in equilibrium
establishes the relation
\begin{eqnarray}
\mu \equiv \mu_n-\mu_p = \mu_e - \mu_{\nu_e} \,, \label{beta}
\end{eqnarray}
allowing the proton chemical potential to be expressed in terms of three
independent chemical potentials: $\mu_n,\mu_e$, and $\mu_{\nu_e}$. At 
densities where $\mu$ exceeds the muon mass, muons can be formed by $e^-
\leftrightarrow \mu^- + \overline\nu_\mu + \nu_e$, hence the muon chemical
potential is 
\begin{eqnarray}
\mu_\mu=\mu_e-\mu_{\nu_e}+\mu_{\nu_\mu}\,, 
\end{eqnarray}
requiring
the specification of an additional chemical potential $\mu_{\nu_\mu}$.
Negatively charged kaons can be formed in the process
$n+e^- \leftrightarrow n + K^-+\nu_e$ when $\mu_{K^-} = \mu$
becomes equal to the energy of the
lowest eigenstate of a $K^-$ in matter.
Weak reactions for the $\Lambda, \Sigma$, and $\Xi$
hyperons are all of the form $B_1+\ell \leftrightarrow B_2+\nu_\ell$, where
$B_1$ and $B_2$ are baryons, $\ell$ is a lepton, and $\nu_\ell$ a neutrino of
the corresponding flavor. The chemical potential for baryon $B$ with baryon
number $b_B$ and electric charge $q_B$ is then given by the general relation
\begin{eqnarray}
\mu_B=b_B\mu_n - q_B\mu \,, \label{echem}
\end{eqnarray}
which leads to 
\begin{equation}
\mu_{\Lambda}= \mu_{\Sigma^0} = \mu_{\Xi^0} = \mu_n \quad;\quad
\mu_{\Sigma^-}= \mu_{\Xi^-} = \mu_n+\mu\quad;\quad
\mu_p = \mu_{\Sigma^+} = \mu_n - \mu\,. \label{murel}
\end{equation}
Applied to matter containing quarks, Eq.~(\ref{echem}) gives
\begin{equation}
\mu_d = \mu_s = (\mu_n+\mu)/3 \quad;\quad
\mu_u = (\mu_n-2\mu)/3 \,.
\end{equation}
If there are no trapped neutrinos present, so that $\mu_{\nu_\ell}=0$, 
there are two 
independent chemical potentials ($\mu_n, \mu_e$) representing conservation
of baryon number and charge.  If trapped neutrinos are present
($\mu_{\nu_\ell}\neq 0$)  further constraints, due to conservation of the
various lepton numbers over the dynamical time scale of evolution, must be
specified.  Defining the concentrations $Y_i=n_i/n$, where the density of
species $i$ is $n_i$ and $n$ is the total {\it baryon} density, the total
(electron type) lepton fraction is $Y_\ell=Y_e+Y_{\nu_e}$.  At the onset of
trapping, during the initial inner core collapse, $Y_\ell\approx 0.4$ and
$Y_e/Y_{\nu_e} \sim 5-7$ depending upon the density \cite{pcollapse}. 
These numbers are not significantly affected by variations in the EOS.  
Following deleptonization, $Y_{\nu_e}=0$, and $Y_e$ can vary widely, both with
density and with the EOS. 
                      
In the case of muons it is generally true that, unless $\mu > m_\mu c^2$, the
net number of $\mu$'s or $\nu_\mu$'s present is zero.  
Because no muon-flavor leptons  are present at the onset
of trapping, $Y_{\nu_\mu}=-Y_\mu$.  Following deleptonization,
$Y_{\nu_\mu}=0$, and $Y_\mu$ is determined by $\mu_\mu=\mu_e$ for $\mu_e >
m_\mu c^2$ and is zero otherwise.  
                                                      
As long as both weak and strong interactions are in equilibrium,  the above
general relationships determine the constituents of the star during its
evolution.  Since electromagnetic interactions give negligible
contributions, it is sufficient to consider the non-interacting (Fermi gas)
forms for the partition  functions of the leptons.  Hadrons, on the other hand,
receive significant contributions at high density from the less well known 
strong interactions.                                                    

To present specific results, we employ a relativistic field theoretical model
in which the baryons, $B$, interact via the exchange of $\sigma$, $\rho$, and 
$\omega$ mesons.  The meson fields are determined by extremization of the
partition function.  The purpose of the $\sigma(\sim550\ 0^+,\ T=0)$ meson is
to simulate the  attractive effect of two pion exchange, while the $\omega(782\
1^-\ T=0)$ provides short range repulsion, and the $\rho(770\ 1^-\ T=1)$
accounts for the isospin dependence of the interaction (the loss of
attraction when the number of neutrons and protons differs). In the case in
which only nucleons are considered, $B=n,p$, this is the well-known Walecka
model \cite{sew}.  The nucleon coupling constants are chosen to reproduce the
binding energy ($\sim 16$ MeV), symmetry energy ($\sim 30-35$ MeV),
equilibrium density ($n_0=0.16\pm 0.01~{\rm fm}^{-3}$), compression modulus
(200 MeV$\lord K_0\lord 300$ MeV), and nucleon Dirac effective mass
$(0.6-0.7)\times 939$ MeV of infinite nuclear matter. The compression modulus
and the effective mass influence the stiffness of the high density EOS and
$M_{max}$. 
                                           
We turn now to the cases in which strange particles are allowed.
In the first model \cite{glenhyp}, we augment the set of baryons $B$ to include
the $\Lambda$, $\Sigma$, and  $\Xi$ hyperons. The hyperon-meson coupling
constants are not well known, but  we can take some guidance from hypernuclear
data.  For example, in nuclear  matter at saturation, the lowest $\Lambda$
level is bound by  28 MeV.  This establishes a correlation between the $\sigma$
and $\omega$ couplings \cite{glenmos}.  Fits to $\Lambda$-hypernuclear levels
constrain the range over which these couplings may be varied to obtain
satisfactory neutron star properties.   The $\rho$-coupling is of less
consequence and may be taken to be of similar order.   The $\Sigma^-$ atom
analysis of Mare\v{s} {\it et al.}  \cite{sigmacou} offers some guidance on
the $\Sigma^-$ couplings, which are of similar magnitude to those of the
$\Lambda$.  Unfortunately, very little is known about the $\Xi$ couplings from
data; hence, we assume them to be of similar magnitude to those of the
$\Lambda$ and $\Sigma^-$.                                                 

In the second model, we allow for kaons in addition to nucleons and leptons.
The kaon-nucleon interaction may also be generated by the exchange of
$\sigma$, $\rho$, and $\omega$ mesons \cite{kpe}.  The qualitative results of 
kaon condensation in this meson exchange model are similar to those of the 
chiral model of Kaplan and Nelson \cite{kapnel}  when the magnitudes of the
kaon-baryon interactions in the two models are  required to yield compatible
kaon optical potentials in nuclear matter.   Friedman {\it et al.} \cite{fgb}
have recently suggested a strongly  attractive $K^-$ optical potential of depth
$-200\pm 20$ MeV for best fits  to kaonic atom data. Theoretically, the major
uncertainty lies in value of the kaon-nucleon sigma term, $\Sigma^{KN}$,  which
depends on the strangeness content of the proton.  For present purposes, we
take $\Sigma^{KN}=344$ MeV on the basis of recent lattice gauge simulations
\cite{dong}.         

Fig. 1 shows the various concentrations as a function of density when the only
hadrons present are nucleons; here, the arrows indicate the central density of
the maximum mass stars. The left hand panel refers to the case in which
the neutrinos have left the star. At high density the proton concentration is
about 30\%, charge neutrality ensuring an equal number of negatively charged
leptons. This relatively large value is the result of the symmetry energy
increasing nearly linearly with density in this model.  Many non-relativistic 
potential models \cite{wff} predict a maximum proton concentration of 10\%. The
effects of neutrino trapping are displayed in the right hand panel. The fact
that $\mu_{\nu_e}\neq0$ in Eq. (\ref{beta}) results in larger values for
$\mu_e$ and $Y_e$. Because of charge neutrality, $Y_p$ is also larger, and it
approaches 40\% at high density.  As is evident from the third column of Table
I,  neutrino-trapping reduces the maximum mass $M_{max}^L$ from the value found
in neutrino-free matter $M_{max}$; although neutrino-trapped nucleons-only
matter contains more leptons and more leptonic pressure, it also contains more
protons and, therefore, less baryonic pressure.  It is also evident from the
table that
thermal effects increase the pressure and therefore the maximum mass, but only
slightly.  Even for $S=2$, the central temperature is only $\sim50$ MeV,
which is 
much less than the nucleon Fermi energies. Thus, in the absence of significant
accretion at late times, a black hole could only form promptly after bounce
from nucleons-only stars, because $M_{max}^L\lord M_{max}$.
                                                                         
Fig. 2 is the analogue of Fig. 1 for the case in which hyperons are allowed to
be present. In the neutrino free case (left-hand panel), the $\Lambda(1116)$
and the $\Sigma^-(1197)$ appear at roughly the same density because the
somewhat higher mass of  the $\Sigma^-$ is compensated by the presence of the
electron chemical potential in the chemical equilibrium condition,
Eq.~(\ref{murel}). More massive, and more positively charged, particles than 
these appear at higher densities.  Following the appearance of the 
negatively charged $\Sigma^-$ hyperon the lepton concentrations fall because
of charge neutrality.  The rapid build-up of the other hyperons with 
increasing density produces a system which is strangeness-rich at high density
and which contains nearly  as many protons as neutrons.  The introduction of
new baryonic species and the lower lepton abundances significantly reduce the
pressure.  This substantially reduces the maximum mass \cite{glenhyp} compared
to the nucleons-only case, as seen from column four of Table I.  
                         
\centerline{Table I} 
\centerline{Maximum neutron star gravitational masses in solar units }
\centerline{ with and without trapped neutrinos}                  
\begin{center}
\begin{tabular}{lcccc}\hline 
&&\multicolumn{3}{c}{Strange hadrons}\\
&$S$ &  None & Hyperons&Kaons\\ \hline
&&&\\[-2mm]
No trapped neutrinos & $0$ & 2.01 & 1.54 & 1.83\\
&&&\\[-2mm]
Trapped neutrinos    & $0$ & 1.94 & 1.77 & 1.93\\
&&&\\[-2mm]
Trapped neutrinos    & $2$ & 1.98 & 1.78 & 1.97\\ \hline
\end{tabular}
\begin{quote}
\noindent The entropy/baryon is denoted by $S$.
\end{quote}
\end{center}

In the case in which neutrinos are trapped (right-hand panel of Fig. 2) the
threshold for the appearance of the $\Sigma^-$ is significantly raised, since   
$\mu=\mu_e -\mu_{\nu_e}$  is much smaller than $\mu_e$ in the untrapped case.
Furthermore, the abundances of all hyperons are smaller, owing to the larger
number of electrons.  These effects stiffen the EOS compared to the neutrino
free case.  Thus, in Table I,  $M_{max}^L$ is $\sim0.2{\rm M}_{\odot}$ larger
than $M_{max}$, exactly the opposite behavior to that obtained in the absence
of strange  particles.                                         
                     
Fig. 3 shows the effect of kaon condensation. We see that in the  neutrino-free
case (left panel) the attractive interaction between $K^-$ mesons and nucleons
allows a condensate to form at a density of 3.6$n_0$. Beyond this density the
proton concentration increases dramatically, balanced by a roughly equal 
number of kaons, and at large densities it becomes almost equal to the neutron
concentration. The EOS is thus softened, and the maximum mass reduced relative
to the nucleons-only case (see column five of Table I), although by a lesser
amount than for hyperons with the models displayed here. 
When neutrinos are trapped (right panel) the $K^-$ threshold is raised and,
given the density for a maximum mass star (arrow), it is clear that the 
condensate plays a much lesser role. Thus the EOS is stiffer and 
the maximum mass is raised. The effect is qualitatively similar to that
engendered by hyperons.  A similar effect is also  obtained if $u,\ d$, and $s$
quarks are allowed to be present \cite{pcl}.

Evolutionary calculations \cite{birth,keiljan} without accretion show that  it
takes on the order of 10--15 seconds for  the trapped neutrino fraction to 
vanish for
a nucleons-only EOS.   In the absence of black hole formation, this evolution
should be qualitatively independent of the EOS \cite{prep}.  This is roughly
borne out in the calculations of Keil and Janka \cite{keiljan}. Fig. 4 shows 
the dependence of the maximum stellar mass upon $Y_{\nu_e}$. When the only
hadrons are nucleons (np) the maximum mass increases with $Y_{\nu_e}$, whereas 
when hyperons (npH) or kaons (npK) are also present, it decreases. Further, the
rate of decrease accelerates for rather small values of $Y_{\nu_e}$ . The
implication is clear. {\it If} ~hyperons, kaons, or other negatively-charged
hadronic species are present, an initially stable star can change into a black
hole after most of the trapped neutrinos have left, and this takes $10-15$ s. 
This happens if the remnant mass $M$ satisfies $M_{max}^L>M>M_{max}$.  
                  
\vskip.75cm
\noindent {\large{3. SUPERNOVA SN1987A} }
\vskip.75cm

\noindent On February 23 of 1987, neutrinos were observed \cite{exp87a} from
the  explosion of supernova SN1987A, indicating that a neutron star, not a
black  hole, was initially present.   (The appearance of a black hole 
would have caused an abrupt cessation of any neutrino signal \cite{bur}.)  The
neutrino signal was observed for a period of at least 12 s, after which
counting statistics fell below measurable limits.  From the handful of events
observed only the average neutrino energy,  $\sim 10$ MeV, and the 
total binding energy release of $\sim (0.1-0.2){\rm M}_\odot$ could be
estimated.  
                                                      
These estimates, however, do not shed much light on the composition of the
neutron star.  This is because, to lowest order, the average neutrino energy
is fixed by the neutrino mean free path in the outer regions of the
protoneutron star. Further, the binding  energy exhibits a  universal
relationship \cite{prep} for a wide class of EOSs, including those with
strangeness  bearing components, namely 
\begin{eqnarray}
B.E. = (0.065\pm 0.01)(M_B/{\rm M}_\odot)^2{\rm M}_\odot \,, 
\end{eqnarray}
where $M_B$ is the baryonic mass. This  allows us only to determine a remnant
gravitational mass of $(1.14-1.55){\rm M}_\odot$, but not the composition.   

The ever-decreasing optical luminosity (light curve) \cite{lcurve}
of the remnant of SN1987A suggests
two arguments against the continued 
presence of a neutron star.  First,  accretion onto a neutron star at 
the Eddington limit is already ruled out for the usual hydrogen-dominated 
Thomson electron scattering opacity.  (However, if the atmosphere surrounding 
the remnant contains a sufficient amount of 
iron-like elements, as Chen and Colgate \cite{cc} suggest, the
appropriate Eddington limit is much lower.) 
Second, a Crab-like pulsar cannot exist in SN1987A, since the
emitted magnetic dipole radiation would be observed in the light curve. 
Either the magnetic field or
the spin rate of the neutron star remnant would have to be much less than in 
the case of the Crab, and what is inferred from other young neutron
stars.  The spin rate of a newly formed
neutron star is expected to be high, however the time scale for the 
generation of a significant magnetic field is not well known and
could be greater than 10 years.  
                                                               
Although most of the binding energy is released during the initial accretion 
and  collapse stage in about 2 seconds after bounce,  the neutrino signal
continued for a period of at least 12 s.   The compositionally-induced changes
in the structure of the star occur on the deleptonization time scale which we
have estimated to be of order 10--15 s \cite{prep}, not on the binding
energy release time scale.   Thus, the duration of the neutrino signal from
SN1987A was  comparable to the time required for the neutrinos initially
trapped in the star to leave.  However,  counting statistics prevented
measurement of a longer duration, and this unfortunate happenstance prevents
one from distinguishing a model in which negatively-charged matter appears and
a black hole forms from a less exotic model in which 
a neutron star still exists.  
As we have pointed out, the maximum stable mass drops by as much as 
$0.2{\rm M}_\odot$ when the trapped neutrinos depart if negatively charged
particles are present,  which could be enough to cause collapse to a black
hole.  
                                                             
Observed neutron stars lie in a very small range of  gravitational masses. The 
smallest range that is consistent with all the data \cite{Taylor,starmass} 
runs from  $1.34{\rm M}_{\odot}$ to $1.44{\rm M}_{\odot}$, the latter value
being the accurate  measurement of PSR1913+16. Thielemann {\it et al.}
\cite{thm} and Bethe and Brown \cite{bb} have estimated the gravitational mass
of the remnant of SN1987A to  be in the range $(1.40-1.56){\rm M}_{\odot}$,
using arguments based on the observed amounts of ejected $^{56}$Ni and/or the
total explosion energy.  This range extends above the largest accurately known
value for a neutron star mass,  $1.44~{\rm M}_\odot$, so the possibility 
exists that the neutron star initially produced in SN1987A could be unstable 
in the cold, deleptonized  state. In this case, SN1987A would have become a 
black hole once it had deleptonized, and no further signal would be expected.
Should this scenario be observationally verified, it would provide strong 
evidence for the appearance of strange matter.

\vskip.75cm
\noindent {\large{4. FUTURE DIRECTIONS}  }
\vskip.75cm

\noindent The emitted neutrinos, of all flavors, are the only direct probe of 
the mechanism of supernova explosions and the structure of newly formed 
neutron stars.  The cooling of the star can yield information on 
the stellar composition for which accurate simulations with appropriate 
neutrino opacities will be necessary. At the same time, further information on 
the crucial question of the strong interactions of strange  particles in dense 
matter is sorely needed --  even near nuclear equilibrium density our 
knowledge is sketchy at present. This emphasizes the need to pin
down the mass shifts of hadrons in dense matter utilizing a reliable many-body
description.
                                
What can be expected in future detections?  In an optimistic scenario, several 
thousand neutrinos from a typical galactic supernova might be seen in  upgraded
neutrino detectors, such as SNO in Canada and Super  Kamiokande in Japan. (For
rough characteristics of present and future neutrino  detectors see
Ref. \cite{bkg}.) The coincidence between the deleptonization time scale and 
the detection time scale that happened in the case of SN1987A will not occur.   
Among the interesting features that could be sought are:   
\begin{enumerate}
\item Possible cessation of a neutrino signal due to black hole formation.
\item Possible burst or light curve feature associated with the onset of
negatively-charged strongly interacting matter near the end of deleptonization,
whether or not a black hole is formed. 
\item Identification of the deleptonization/cooling epochs by changes in
luminosity evolution or neutrino flavor distribution.
\item Determination of a radius-mean free path correlation from the luminosity
decay time or the onset of neutrino transparency.
\item Determination of the neutron star mass from the universal  binding
energy-mass relation.
\end{enumerate}
To realize the above goals and to adequately discriminate between the various
possibilities, detailed information about the characteristics of neutrino
detectors must be made available. This is especially important in deciphering
the time evolution of the neutrino signal, even if a large number of neutrinos
are detected.
\newpage
{\small
\begin{flushleft}
{\bf Acknowledgements}
\end{flushleft}
\noindent This work was supported by NASA grant NAG52863 and by the
U.S. Dept. of Energy  under grants DOE/DE-FG02-88ER-40388,
DOE/DE-FG02-87ER-40328 and DOE/DE-FG02-87ER-40317.  }

\begin{flushright}
PAUL J. ELLIS\\[-2mm]
{\it School of Physics and Astronomy}\\[-2mm]
{\it University of Minnesota}\\[-2mm]
{\it Minneapolis, MN 55455}\\
\vskip1cm
JAMES M. LATTIMER\\[-2mm]
{\it Department of Earth and Space Sciences}\\[-2mm]
{\it State University of New York}\\[-2mm]
{\it Stony Brook, NY 11794}\\
\vskip1cm
MADAPPA PRAKASH\\[-2mm]
{\it Department of Physics}\\[-2mm]
{\it State University of New York}\\[-2mm]
{\it Stony Brook, NY 11794}\\
\end{flushright}
\newpage
\renewcommand{\baselinestretch}{1}
\newpage
\renewcommand{\baselinestretch}{1.5}
\centerline{\bf Figure Captions}

\noindent Fig. 1.  Individual concentrations, $Y_i$, as a function of the
baryon density ratio $u=n/n_0$, where $n_0$ is the density of 
equilibrium nuclear matter. The arrows indicate the central density of the
maximum mass stars. Here nucleons, electrons, and muons are in beta 
equilibrium at an entropy per baryon $S=1$. Left panel: neutrino free. Right 
panel: with trapped neutrinos ($Y_{\ell}=0.4$). \hfill \\             

\noindent Fig. 2. As for Fig.1, but for matter which contains hyperons,
as well as nucleons, electrons, and muons. \hfill \\             

\noindent Fig. 3. As for Fig.1, but for matter which contains a kaon 
condensate, as well as nucleons, electrons, and muons. \hfill \\             

\noindent Fig. 4. Maximum neutron star mass as a function of
$Y_{\nu_e}$ for hadronic matter with only nucleons (np) or with nucleons
and hyperons (npH) or kaons (npK). \hfill \\             

\newpage

\begin{figure}[p]
\begin{center}
\leavevmode
\epsfxsize = 6in
\epsfysize = 5in
\epsffile{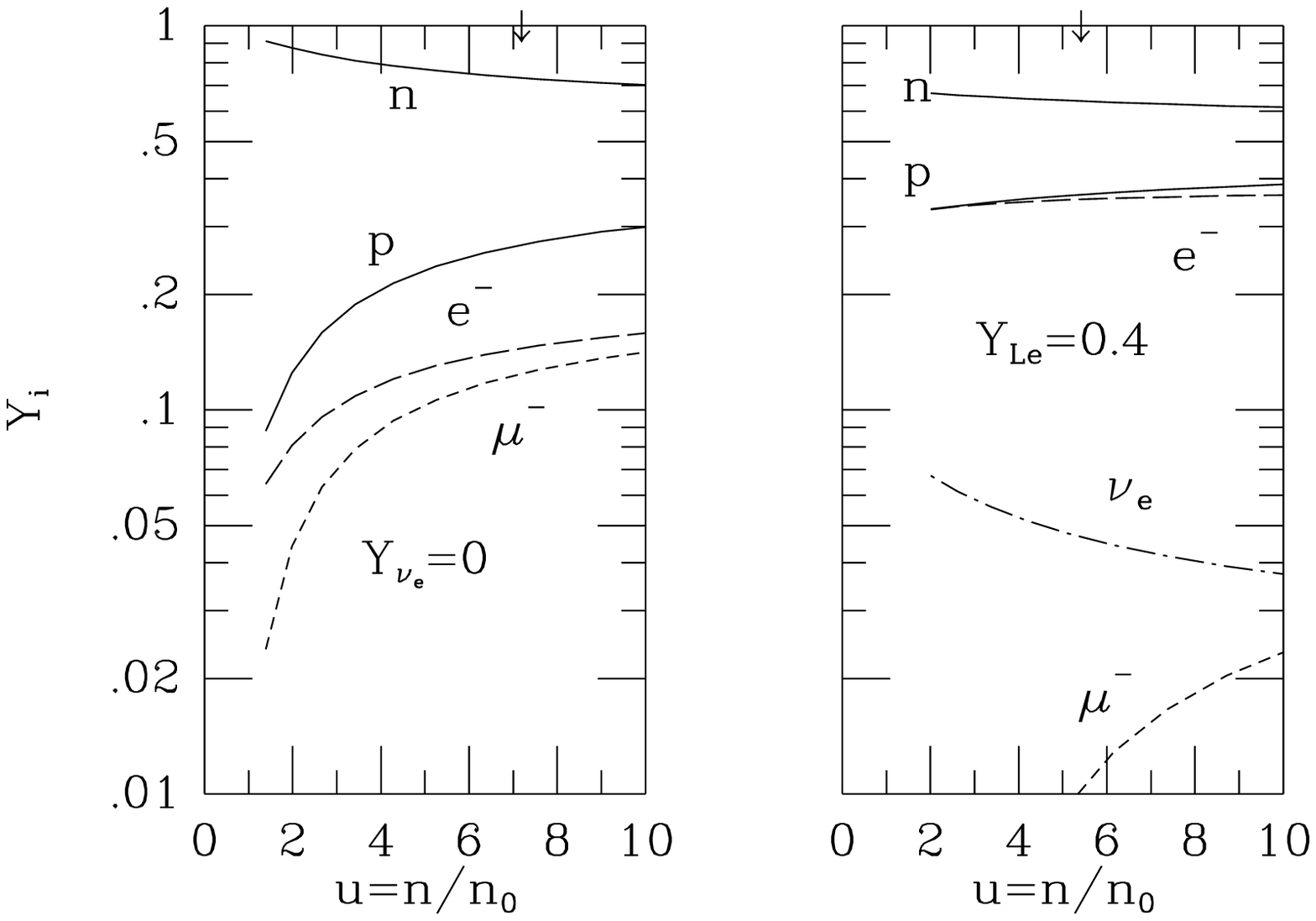}
\vspace*{0.15in} 
\caption{} 
\end{center}
\end{figure}

\begin{figure}[p]
\begin{center}
\leavevmode
\epsfxsize = 6in
\epsfysize = 5in
\epsffile{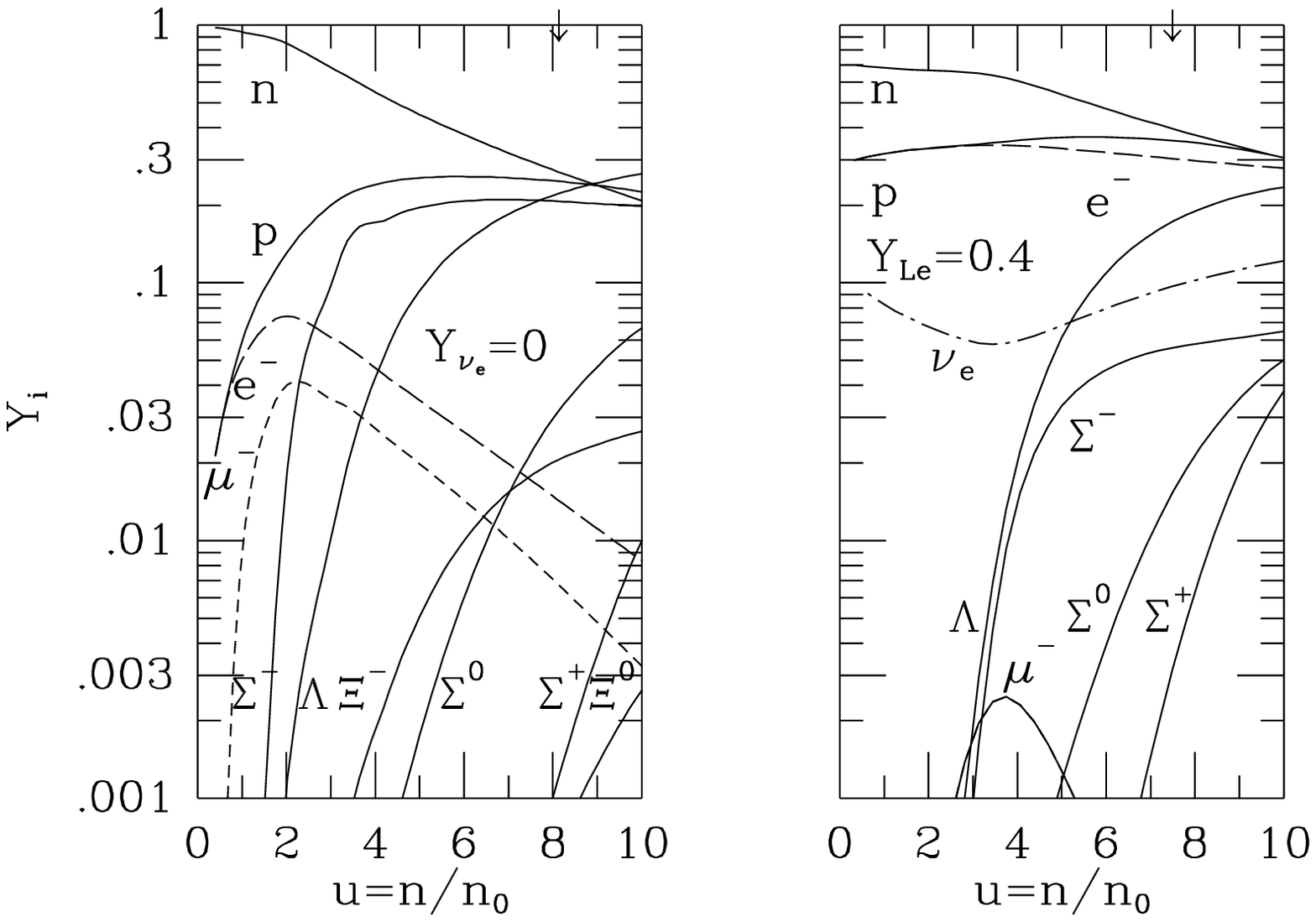}
\vspace*{0.15in} 
\caption{} 
\end{center}
\end{figure}

\begin{figure}[p]
\begin{center}
\leavevmode
\epsfxsize = 6in
\epsfysize = 5in
\epsffile{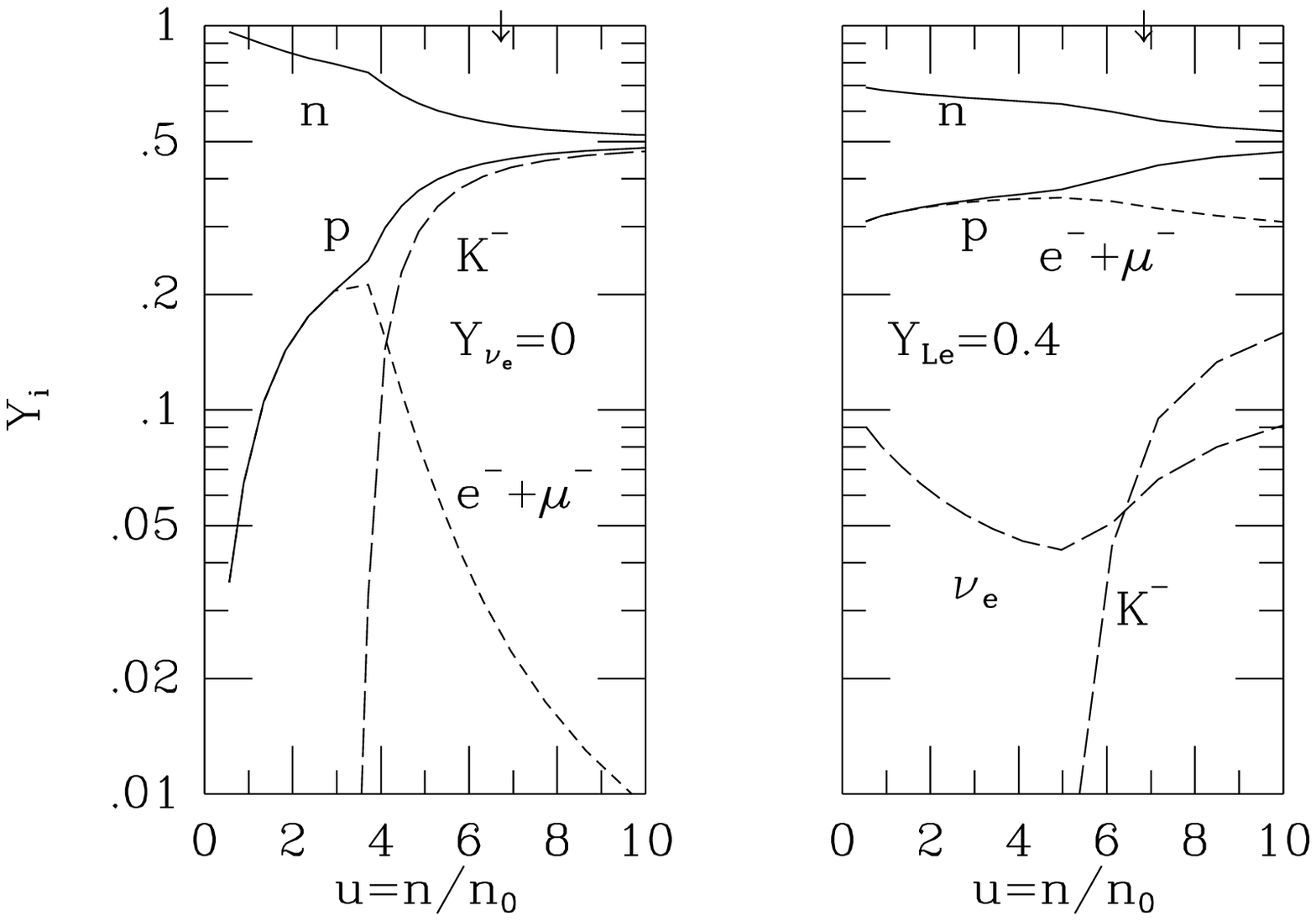}
\vspace*{0.15in} 
\caption{} 
\end{center}
\end{figure}

\begin{figure}[p]
\begin{center}
\leavevmode
\epsfxsize = 6in
\epsfysize = 6in
\epsffile{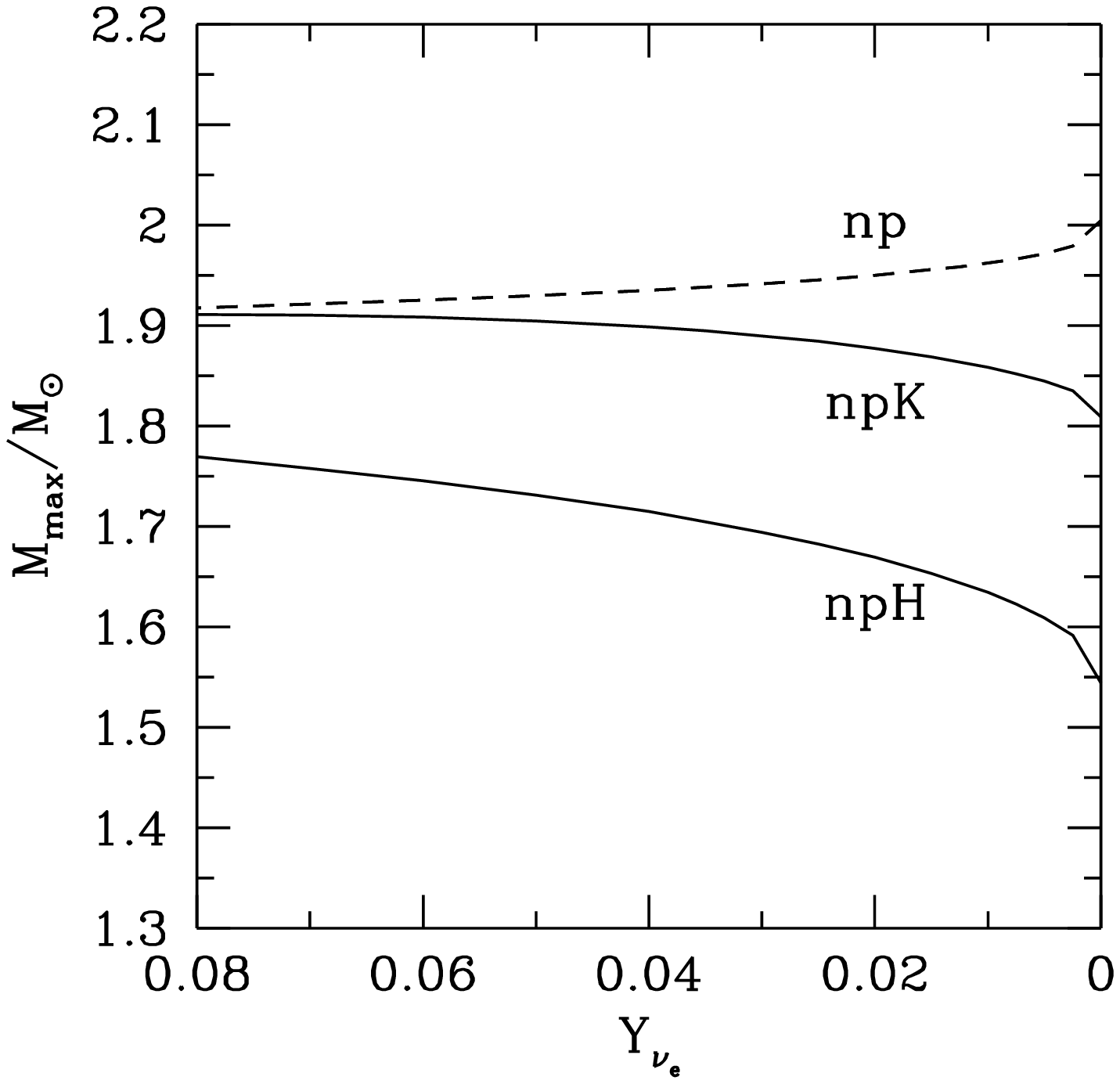}
\vspace*{0.15in} 
\caption{} 
\end{center}
\end{figure}

\end{document}